# A novel approach for the diagnosis of ventricular tachycardia based on phase space reconstruction of ECG

Koulaouzidis G[1], Das S[2,] Cappiello G[2], Mazomenos EB[2], Maharatna K[2], Morgan J[1]

[1] University Hospital Southampton NHS Foundation Trust, Southampton, UK

[2] School of Electronics and Computer Science, University of Southampton, UK

Corresponding Author

Koulaouzidis George

University Hospital Southampton NHS Foundation Trust, Southampton, UK

geokoul@hotmail.com

Ventricular arrhythmias comprise a group of disorders which manifest clinically in a variety of ways from ventricular premature beats (VPB) and no sustained ventricular tachycardia (in healthy subjects) to sudden cardiac death due to ventricular tachyarrhythmia in patients with and/or without structural heart disease.

Ventricular fibrillation (VF) and ventricular tachycardia (VT) are the most common electrical mechanisms for cardiac arrest. Accurate and automatic recognition of these arrhythmias from electrocardiography (ECG) is a crucial task for medical professionals. The purpose of this research is to develop a new index for the differential diagnosis of normal sinus rhythm (SR) and ventricular arrhythmias, based on phase space reconstruction (PSR).

PSR can map a time-series to a phase space trajectory in multi-dimensional space using a time delay embedding technique. PSR is a technique widely used to observe nonlinear behaviour of dynamical system and detect such small desynchronisation phenomena, which is difficult to identified, by simple observation (1-3). Over the last decade, PSR of ECG has been successfully used in a number of heart disease detection related applications (4-7).

For the analysis we used 32 ECGs with sinus rhythm from the PTB Diagnostic ECG database and 32 ECGs from subjects with VT/VF from the Creighton University Ventricular Tachyarrhythmia Database. First, a reconstruction of phase portrait was performed with the method of delays. In this method, we insert a delay "$\tau$" in the original time-series X(t) which produces a delayed version of X(t), the Y(t)=X(t-$\tau$). Afterwards, the phase-space diagram is reconstructed by plotting Y(t) against X(t). Thereafter, the method of box counting was applied to analyse the behaviour of the phase trajectories, e.g. the number of trajectories and their spread. In this technique, the entire phase-space diagram is represented as an image of N×N pixel, where N is an integer (Fig 1). The pixels through which at least one trajectory has passed are considered as black boxes ($n_b$) and the others are considered as white boxes ($n_w$). The degree of complexity of the phase-space portraits is described using a metric, defined as the ratio of the number of black boxes ($n_b$) or pixels visited by the trajectories and the total number of pixels ($n_w+n_b=N^2$).

In this study, a different approach is suggested. Run-time statistical measures like mean ($\mu$), standard deviation ($\sigma$) and coefficient of variation ($CV = \sigma/\mu$), skewness ($\gamma$), kurtosis ($\beta$) for the box counting in phase-space diagrams are studied for a sliding window of 10 beats of ECG signal.

When the subjects are in SR the phase portraits show regular structure, while during the VT/VF period the phase portraits show chaotic motion (Fig 1). In the 32 health subjects with SR, the analysis showed that $\mu$ and $\sigma$ trends are almost uniform throughout the time. On the other hand, in the arrhythmic subjects both $\mu$ and $\sigma$ showed sudden increase at the stage of VT/VF onset. In order to identify accurately the onset of arrhythmia, $CV = \sigma/\mu$ trends was introduced. $CV$ is always bounded within an upper limit of $CV<0.05$, having values between $>0$ and $<0.05$. On the other hand, the arrhythmic subjects showed increase in $CV$, an increase that correlates with the onset of arrhythmia. During the arrhythmia the $CV$ remained stable above the value of $>0.05$. It is worth to mention that after the appearance of a VPB the value of $CV$ normalized after an initial increase. Therefore, the upper threshold was considered for the healthy subjects $CV_{th}=0.05$. Similar pattern was observed also with the kurtosis, in which despite the inter-person variability in each case, the kurtosis crossed an upper limit of $\beta_{th}<6$ several times which was considered as the cut-off point between subjects with SR and VT/VF. In order to optimise the accuracy of our diagnosis a new index ($J$) was proposed as a combination of the trends $\beta$ and $CV$.

$$J = w\frac{CV}{CV_{th}} + (1-w)\frac{\beta}{\beta_{th}}$$

The upper normal limit of index $J$ is the value of $J_{th}=1$, while the crossing of upper bounds $CV_{th}$ and $\beta_{th}$ will be reflected in crossing of the threshold for the index of $J_{th}=1$. In the above equation the weight $w$ keep the balance in the impact of $CV$ and $\beta$ trends in index $J$. So in healthy subjects with equal weights on the two parts the $w$ would be 0.5. While in the arrhythmic subjects, it is observed that the trends cross the critical threshold of $J=1$ at different time instants. For $w=0$ the full emphasis is on the kurtosis and with gradual increase of $w$, the impact of $CV$ increases slowly and consequently impact of kurtosis decreases. With $w=1$, the prediction index simply represented the $CV$ trend. In this way the chance of misdiagnosis of arrhythmia has been minimized.

In conclusion, we propose a novel statistical index for diagnosis of ventricular arrhythmia, using the phase-space reconstruction method of long-term ECG time-series. We found that two thresholds $CV<0.05$ and kurtosis $\beta<6$ and the mixture of these in the proposed index $J<1$ identified the SR from VA. Therefore, the index can be a beneficial clinical tool especially for physicians, general practitioners and medical staff with limited expertise in cardiology.

**Fig 1.** Black and white image of a window of ten beats for (a) normal looking ECG before arrhythmia (b) with VPB before arrhythmia, (c) VT, (d) VF.

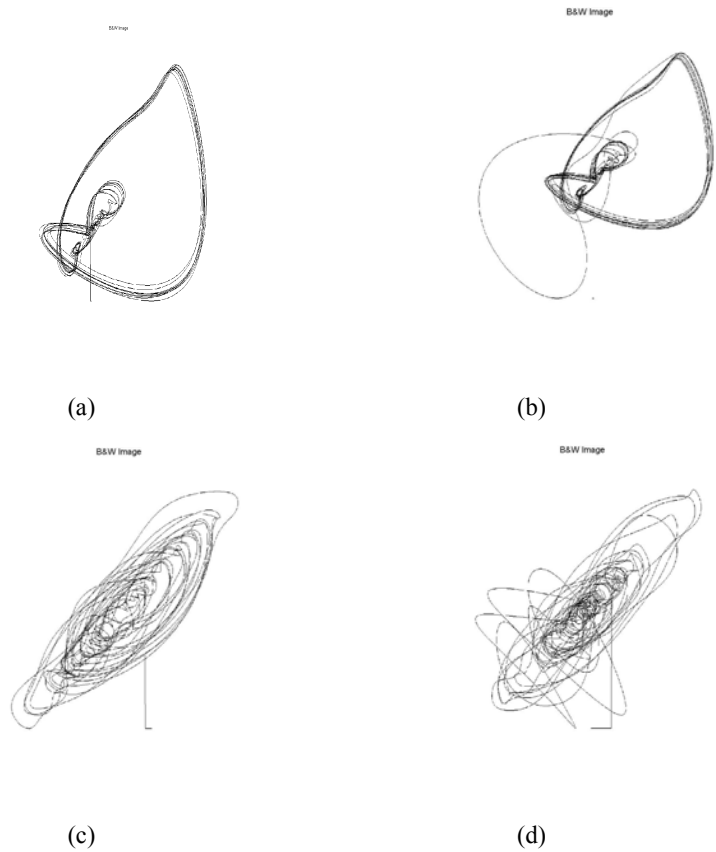

(a)     (b)

(c)     (d)

**Fig 2.** Box counting mean and standard deviation trends for patient with healthy and arrhythmic patient respectively.

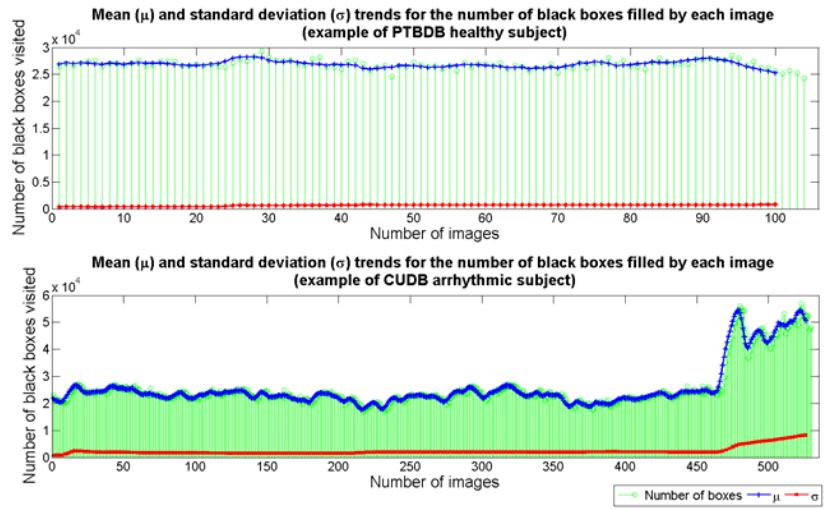